\documentclass[12pt, epsfig]{article}
\usepackage{epsfig}
\usepackage{slashed}

\begin{document}
\newcommand{\beq}{\begin{equation}}
\newcommand{\eeq}{\end{equation}}
\newcommand{\bea}{\begin{eqnarray}}
\newcommand{\eea}{\end{eqnarray}}
\newcommand{\beas}{\begin{eqnarray*}}
\newcommand{\eeas}{\end{eqnarray*}}
\newcommand{\defi}{\stackrel{\rm def}{=}}
\newcommand{\non}{\nonumber}
\newcommand{\bquo}{\begin{quote}}
\newcommand{\enqu}{\end{quote}}
\newcommand{\p}{\partial}
\def\de{\partial}
\def\Tr{ \hbox{\rm Tr}}
\def\const{\hbox {\rm const.}}
\def\o{\over}
\def\im{\hbox{\rm Im}}
\def\re{\hbox{\rm Re}}
\def\bra{\langle}\def\ket{\rangle}
\def\Arg{\hbox {\rm Arg}}
\def\Re{\hbox {\rm Re}}
\def\Im{\hbox {\rm Im}}
\def\diag{\hbox{\rm diag}}

\def\stroke{\vrule height8pt width0.4pt depth-0.1pt}
\def\topfleck{\vrule height8pt width0.5pt depth-5.9pt}
\def\botfleck{\vrule height2pt width0.5pt depth0.1pt}
\def\Zmath{\vcenter{\hbox{\numbers\rlap{\rlap{Z}\kern 0.8pt\topfleck}\kern
2.2pt
                   \rlap Z\kern 6pt\botfleck\kern 1pt}}}
\def\Qmath{\vcenter{\hbox{\upright\rlap{\rlap{Q}\kern
                   3.8pt\stroke}\phantom{Q}}}}
\def\Nmath{\vcenter{\hbox{\upright\rlap{I}\kern 1.7pt N}}}
\def\Cmath{\vcenter{\hbox{\upright\rlap{\rlap{C}\kern
                   3.8pt\stroke}\phantom{C}}}}
\def\Rmath{\vcenter{\hbox{\upright\rlap{I}\kern 1.7pt R}}}
\def\Z{\ifmmode\Zmath\else$\Zmath$\fi}
\def\Q{\ifmmode\Qmath\else$\Qmath$\fi}
\def\N{\ifmmode\Nmath\else$\Nmath$\fi}
\def\C{\ifmmode\Cmath\else$\Cmath$\fi}
\def\R{\ifmmode\Rmath\else$\Rmath$\fi}
\def\QATOPD#1#2#3#4{{#3 \atopwithdelims#1#2 #4}}
\def\stackunder#1#2{\mathrel{\mathop{#2}\limits_{#1}}}
\def\stackreb#1#2{\mathrel{\mathop{#2}\limits_{#1}}}
\def\Tr{{\rm Tr}}
\def\res{{\rm res}}
\def\Bf#1{\mbox{\boldmath $#1$}}
\def\balpha{{\Bf\alpha}}
\def\bbeta{{\Bf\beta}}
\def\bgamma{{\Bf\gamma}}
\def\bnu{{\Bf\nu}}
\def\bmu{{\Bf\mu}}
\def\bphi{{\Bf\phi}}
\def\bPhi{{\Bf\Phi}}
\def\bomega{{\Bf\omega}}
\def\blambda{{\Bf\lambda}}
\def\brho{{\Bf\rho}}
\def\bsigma{{\bfit\sigma}}
\def\bxi{{\Bf\xi}}
\def\bbeta{{\Bf\eta}}
\def\d{\partial}
\def\der#1#2{\frac{\d{#1}}{\d{#2}}}
\def\Im{{\rm Im}}
\def\Re{{\rm Re}}
\def\rank{{\rm rank}}
\def\diag{{\rm diag}}
\def\2{{1\over 2}}
\def\ntwo{${\cal N}=2\;$}
\def\4N{${\cal N}=4$}
\def\none{${\cal N}=1\;$}
\def\x{\stackrel{\otimes}{,}}
\def\beq{\begin{equation}}
\def\eeq{\end{equation}}
\def\ba{\beq\new\begin{array}{c}}
\def\ea{\end{array}\eeq}
\def\be{\ba}
\def\ee{\ea}
\def\stackreb#1#2{\mathrel{\mathop{#2}\limits_{#1}}}
\def\baselinestretch{1.0}

\begin{titlepage}

\begin{flushright}
September 9, 2010
\end{flushright}

\vspace{1mm}

\begin{center}
{\large  {\bf The sparticle spectrum   \\[4mm]
in Minimal gaugino-Gauge Mediation }}
\end{center}

\vspace{1.0mm}

\begin{center}
{\large   {\sc Roberto Auzzi}$^{(1)}$} and {\large  {\sc Amit Giveon}$^{(2)}$}

\vspace{10.0mm}

{\it \footnotesize
Racah Institute of Physics, The Hebrew University,  \\ Jerusalem 91904, Israel}
\\[4mm]
$^{(1)}$ {\tt auzzi@phys.huji.ac.il} \\
$^{(2)}$ {\tt giveon@phys.huji.ac.il}
\end{center}

\vspace{10.0mm}

\begin{abstract}

We compute the sparticle mass spectrum in the minimal
four-dimensional construction that interpolates between
gaugino mediation and ordinary gauge mediation.

\end{abstract}

\end{titlepage}

\vfill\eject

\section{Introduction}

\begin{figure}[h]
 \centerline{\includegraphics[width=3in]{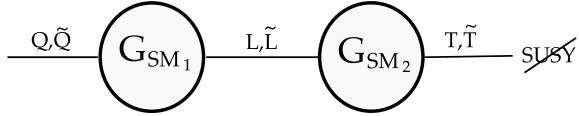}}
 \caption{\footnotesize Quiver diagram for our setting. }
\label{quiver}
\end{figure}

In this note, we compute the soft masses in the
minimal four-dimensional construction \cite{Cheng,Csaki}
of ``gaugino mediation'' \cite{Kaplan:1999ac,Chacko:1999mi}
(see also \cite{gkk} for a recent discussion).
The model is presented in figure \ref{quiver}.
The chiral superfields $Q,\tilde Q$ are the matter fields of MSSM,
$L,\tilde L$ is a single pair of ``link fields'' in the bifundamental
of $G_{SM_1}\times G_{SM_2}$, whose VEV breaks this product group
to the diagonal Standard-Model (SM) gauge group,
$G_{SM}$, and $T,\tilde T$ is a single pair of messengers,
which couple to the spurion of SUSY-breaking, $S$,
whose scalar components get VEVs,
\beq
S=M+\theta^2 F \, , \label{sss}
\eeq
as in Minimal Gauge Mediation (MGM).
The superpotential of the model takes the form
\beq W= S T \tilde{T} + K (L \tilde{L} -v^2) \, , \eeq
where $K$ is a Lagrange multiplier superfield, introduced to set the VEV of the link fields to $v$.

For simplicity, we first take $G_{SM_1}\times G_{SM_2}$ to be
$U(1)_1\times U(1)_2$.~\footnote{The generalization to $G_{SM}=SU(3)\times SU(2)\times U(1)$
is simple, and will be presented in section \ref{mssm-section}.}
Let us introduce the two dimensionless parameters $x$ and $y$:
\beq
x\equiv{F\over M^2}~,\qquad
y\equiv{m_v\over M}~,\qquad m_v\equiv 2v\sqrt{g_1^2+g_2^2} \, ,  \label{xxyy}
\eeq
where $m_v$ is the mass of the massive combination of
gauge bosons of the broken $U(1)_1\times U(1)_2$,
and $g_{1,2}$ are the gauge couplings of $U(1)_{1,2}$, respectively.
The parameter $x$ is a measure of the SUSY-breaking scale, $F/M$,
relative to the messenger scale, $M$,
while $y$ interpolates between MGM (as $y\to\infty$) and minimal gaugino mediation (when $y\ll 1$).
We thus refer to this model as ``Minimal gaugino-Gauge Mediation'' (MgGM).

The main result of this note is the following.
The soft scalar masses (at the messenger scale) in this theory, $m^2_{\tilde f}$, are obtained
by adding to the two-loop integrands in \cite{Martin1996} the common factor,
\beq
f(k^2,m_v^2) \equiv \left({m_v^2\over k^2-m_v^2}\right)^2~, \label{factor}
\eeq
namely,
\beq
m^2_{\tilde f}=\int\frac{d^4 p}{(2 \pi) ^4}\frac{d^4 k}{(2 \pi)^4}  \left(\cite{Martin1996}\right)f~,
\label{mmff}
\eeq
where (\cite{Martin1996}) in the integrand is the same as for MGM in \cite{Martin1996},
and the momentum $k$ amounts to the one on the massless propagator in each of the two-loop diagrams.
The SM coupling, $g_e$, is given in terms of $g_{1,2}$ by
\beq
{1\over g_{e}^2}={1\over g_1^2}+{1\over g_2^2}\, . \label{gsm}
\eeq
When $v$ is much smaller than $M$,
eq. (\ref{mmff}) implies that $m^2_{\tilde f}$ has a suppression factor of order $v^2/M^2$
relative to MGM, while $f\to 1$ if $v\to\infty$, in which case one recovers the results
of MGM \cite{Dimopoulos:1996gy,Martin1996}. On the other hand,
the gaugino masses, $m_{\tilde g}$, are as in MGM, with the SM coupling
given by (\ref{gsm}), for any $v$.

This note is organized as follows. In section  \ref{setting-section},
the theoretical setting is introduced.
In section \ref{graphs-section},
the two-loop graphs contributing to the sfermions mass are discussed,
and in section \ref{integrals-section} they are evaluated.
In section \ref{mssm-section}, we present the soft masses for MSSM,
and in section \ref{conclusion-section}  we discuss our results.
Finally, in a couple of appendices, we list some
technical details about the evaluation
of the gaugino and sfermion graphs.

\section{Theoretical setting}
\label{setting-section}

We consider the setting in figure \ref{quiver}, discussed in the introduction.
The matter fields are taken with charge $\pm1$,
with the following sign choice:
\beq D_\mu Q=\partial_\mu Q + i g_1 A^1_\mu Q   \, , \qquad   D_\mu T=\partial_\mu T + i g_2 A^2_\mu T   \, ,    \eeq
\[  D_\mu  L   =\partial_\mu L + i g_1 A^1_\mu L  - i g_2   A_\mu^2  L \, .\]
The potential is the sum of D and F terms:
\beq
V_D=\frac{g_1^2}{2} \left( Q Q^\dagger -\tilde{Q}^\dagger  \tilde{Q} +L L^\dagger -\tilde{L}^\dagger  \tilde{L} \right)^2
+ \frac{g_2^2}{2}   \left( -L L^\dagger +\tilde{L}^\dagger  \tilde{L} +T T^\dagger -\tilde{T}^\dagger  \tilde{T} \right)^2 \, ,
\eeq
\[
V_F= |L \tilde{L} -v^2|^2 + |K L|^2 + |K \tilde{L}|^2 +
  |S T|^2 + |S \tilde{T}|^2 +|T \tilde{T} + F|^2 \, .
\]
The VEVs of the scalars are:
\[ L=\tilde{L}=v \, ,\qquad K=T=\tilde{T}=Q=\tilde{Q}=0 \, , \qquad  S=M \, .\]

\subsection{Tree-level masses}
After the VEV insertion, the following  term gives mass to a combination of
the two $U(1)$'s:
\beq 2 v^2  (g_1^2+g_2^2) \, \left( \frac{g_1 A_\mu^1 - g_2 A_\mu^2}{\sqrt{g_1^2+g_2^2}}   \right)^2 \, ,\eeq
which gives $m_v=2 v \sqrt{g_1^2+g_2^2}$ for the combination of the two vectors which gets a mass.

The part of the Lagrangian corresponding to the scalar masses reads:
\beq
\left(\begin{array}{cc}
\delta T^* & \delta \tilde{T}
\end{array}\right) \,
 \left(\begin{array}{cc}
 M^2  &  F  \\
 F  & M^2   \\
\end{array}\right) \,
 \left(\begin{array}{c}
\delta T  \\
\delta \tilde{T}^* \\
\end{array}\right)  + 2  v^2 |\delta K|^2
\eeq
\[+ v^2  |\delta L +\delta \tilde{L}|^2
+\frac{g_1^2 + g_2^2}{2} v^2 (\delta L +\delta L^*-\delta \tilde{L} -\delta \tilde{L}^*)^2 \, .
\]
The imaginary part of the scalar  $\frac{(\delta L - \delta \tilde {L})}{\sqrt{2}}$ is eaten by Higgs mechanism;
the real part of the same scalar takes the same mass $m_v$ as the gauge boson
(it is in the same supermultiplet).
The scalar messengers $T_{\pm}=(T \pm \tilde{T}^*)/\sqrt{2}$ get mass squared
$ m_{\pm}^2 = M^2 \pm  F $.

The piece corresponding to the fermion masses is:
\beq
iv \sqrt{2}  ( g_1 \lambda_1 -g_2 \lambda_2  )(\psi_L - \psi_{\tilde{L}})
 - (M \psi_T \psi_{\tilde{T}}  + v \psi_K \psi_{\tilde{L}}
+ v \psi_K \psi_L )  + {\rm c. c.}  \eeq
The  combination
\beq \lambda_A=i \, \frac{g_2 \lambda_1 + g_1 \lambda_2}{\sqrt{g_1^2+g_2^2}} \, ,   \label{l1}\eeq
 remains massless at tree level, while
\beq
\lambda_B= i \,  \frac{g_1 \lambda_1 - g_2 \lambda_2}{\sqrt{g_1^2+g_2^2}} \, , \label{l2} \qquad
\eta=\frac{\psi_{\tilde{L}}-\psi_L}{\sqrt{2}} \, ,
\eeq
mix to make the following  Dirac fermion
\beq
\kappa=  \left(\begin{array}{c}
 (\lambda_B)_\alpha \\
(\eta^*  )^{\dot{\alpha}} \\
\end{array}\right) \, , \label{kkkk}
\eeq
whose mass is $m_v$.
Finally, the fermionic messengers $\psi_T, \psi_{\tilde{T}}$ get a mass $m_f=M$.

\subsection{Gaugino couplings}
In Weyl spinor notation, the gaugino couplings with the $Q,\tilde{Q},T,\tilde{T}$ hypermultiplets are:
\beq
- i g_2 \sqrt{2} \left( T \psi_T^*  \lambda^*_2 -T^* \psi_{T} \lambda_2
- \tilde{T} \psi^*_{\tilde{T}} \lambda^*_2 + \tilde{T}^* \psi_{\tilde{T}} \lambda_2
\right)  \eeq
\[
- i g_1\sqrt{2} \left( Q \psi_Q^*  \lambda^*_1 -Q^* \psi_{Q} \lambda_1
- \tilde{Q} \psi^*_{\tilde{Q}} \lambda^*_1 + \tilde{Q}^* \psi_{\tilde{Q}} \lambda_1
\right) \, .
\]
After some manipulations these couplings are:
\beq
 \frac{1}{ \sqrt{g_1^2+g_2^2}}
 \left( g_1 g_2 T_+ (\psi_T^* \lambda_A^* -\psi_{\tilde{T}} \lambda_A)
 + g_1 g_2 T_- (\psi_T^* \lambda_A^* +\psi_{\tilde{T}} \lambda_A)
 \right. \eeq
 \[ \left.
 +g_2^2 T_+ (\psi_{\tilde{T}} \lambda_B -\psi_T^* \lambda_B^*)
 -g_2^2 T_-  (\psi_T^* \lambda_B^* + \psi_{\tilde{T}} \lambda_B)
\right)
\]
\[
+\frac{\sqrt{2}}{ \sqrt{g_1^2+g_2^2}}  Q \left( g_1 g_2    \psi_Q^* \lambda_A^* + g_1^2 \psi_Q^* \lambda_B^* \right) + {\rm c. c.} \, .
\]
It is useful to write some of these couplings in Dirac notation;
 the following spinors are introduced for this purpose:
\beq \omega_T =  \left(\begin{array}{c}
(\psi_T)_\alpha  \\
(\psi_{\tilde{T}}^*  )^{\dot{\alpha}} \\
\end{array}\right)  \, , \qquad
\omega_{Q} =  \left(\begin{array}{c}
(\psi_{Q})_\alpha  \\
(\psi_{\tilde{Q}}^*  )^{\dot{\alpha}} \\
\end{array}\right)  \, , \qquad
\lambda_M=   \left(\begin{array}{c}
(\lambda_A)_\alpha  \\
(\lambda_A^*  )^{\dot{\alpha}} \\
\end{array}\right)  \, .
\eeq
The couplings involving $\lambda_{M}$, $\omega_{Q,T}$
and $Q,T_+,T_-$ are:
\beq
g_e \left(   \sqrt{2}   Q
\bar{\omega}_{Q}  \frac{1+\gamma^5}{2} \lambda_M+
 T_+ \bar{\omega}_T \gamma_5  \lambda_M
+  T_- \bar{\omega}_T  \lambda_M \right)+
\rm{c.c.}~,
\eeq
where $g_e$ is defined in (\ref{gsm}),
while the couplings involving the Dirac spinor $\kappa$ (\ref{kkkk}) are:
\beq   \frac{1 }{\sqrt{g_1^2+g_2^2} }
\left(   \sqrt{2}   g_1^2  Q
\bar{\omega}_Q \frac{1+\gamma^5}{2} \kappa^c  +
  g_2^2 T_+
\left(    \bar{\omega}_T  \frac{1-\gamma_5}{2} \kappa -
\bar{\omega}_T  \frac{1+\gamma_5}{2}  \kappa^c \right) \right.
\eeq
\[ \left.
 -  g_2^2  T_-
\left(    \bar{\omega}_T  \frac{1-\gamma_5}{2} \kappa +
\bar{\omega}_T  \frac{1+\gamma_5}{2}  \kappa^c \right)
\right)+ {\rm c.c.}  \, .
\]
Here $\kappa^c$ is the charge conjugate spinor of $\kappa$:
\beq \kappa^c = \left(\begin{array}{c}
 (\eta)_\alpha \\
(\lambda_B^*  )^{\dot{\alpha}} \\
\end{array}\right) \, .\eeq

\section{Calculation of the sfermion masses}
\label{graphs-section}

The aim is to generalize the two-loops calculations by Martin \cite{Martin1996} in
minimal gauge mediation. These graphs come in three different classes:
there is a graph due to the exchange of scalars,
some graphs which are due to the exchange of gauge bosons
and a graph which is due to exchange of gauginos.
In this section we examine each of these contributions separately.

\subsection{Scalar  graph}

\begin{figure}[h]
 \centerline{\includegraphics[width=3in]{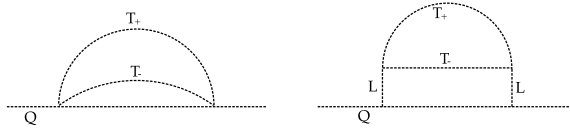}}
 \caption{\footnotesize Graph corresponding to the contribution due
 to D-term (on the left at infinite $v$, on the right at finite $v$).  }
\label{graph7}
\end{figure}

The graph  corresponding to the contribution due to scalar exchange  is shown in figure \ref{graph7}.
The two $\Phi^4$ interactions
\[ g_e^2 Q Q^* (T_+ T_-^* + T_+^* T_-) \, , \]
of the minimal gauge mediation case
  are  replaced by four
$\Phi^3$ interactions.

The detailed form of these interactions is:
\beq
-2 v \left( g_1^2  \frac{\delta L_R -   \delta \tilde{L}_R}{\sqrt{2}} Q Q^*
+ g_2^2  \frac{ \delta L_R - \delta  \tilde{L}_R}{\sqrt{2}} (T_+ T_-^* +T_- T_+^*) \right)\, ,
\eeq
where $L=v+(\delta L_R + i \delta L_I)/\sqrt{2}$  and $\tilde{L}=v+(\delta \tilde{L}_R + i \delta \tilde{L}_I)/\sqrt{2}$.
 Notice that this cubic vertex couples just with the eigenvector of the mass matrix
 whose mass is $m_v$ (\ref{xxyy}).
 So the propagator that must be inserted between each couple of vertical cubic vertices
 is   \[ \frac{i}{k^2-m_v^2} \, .\]
 In the $v \rightarrow \infty$ limit the
usual $\Phi^4$ interaction is recovered, with the diagonal $U(1)$ effective coupling constant $g_e^2$.
A direct evaluation gives:
\beq -2
 \int \frac{d^4 p}{(2 \pi) ^4}  \frac{d^4 k}{(2 \pi)^4}
 \frac{1}{(k-p)^2-m_+^2} \frac{1}{p^2-m_-^2} \frac{1}{k^2}  \left(\frac{4 g_1^2 g_2^2 v^2}{k^2-m_v^2}\right)^2
\eeq
\[ = -2g_e^4
 \int \frac{d^4 p}{(2 \pi) ^4}  \frac{d^4 k}{(2 \pi)^4}
 \frac{1}{(k-p)^2-m_+^2} \frac{1}{p^2-m_-^2} \frac{1}{k^2}
 f(k^2,m_v^2)
 \]
\[ =
\int\frac{d^4 p}{(2 \pi) ^4}\frac{d^4 k}{(2 \pi)^4}  \left(\cite{Martin1996}\right)f~,
\]
where $f$ is given in (\ref{factor}).
This proves the claim in eq. (\ref{mmff}) for the scalar graph.

\subsection{Gauge boson graphs}

\begin{figure}[h]
 \centerline{\includegraphics[width=3in]{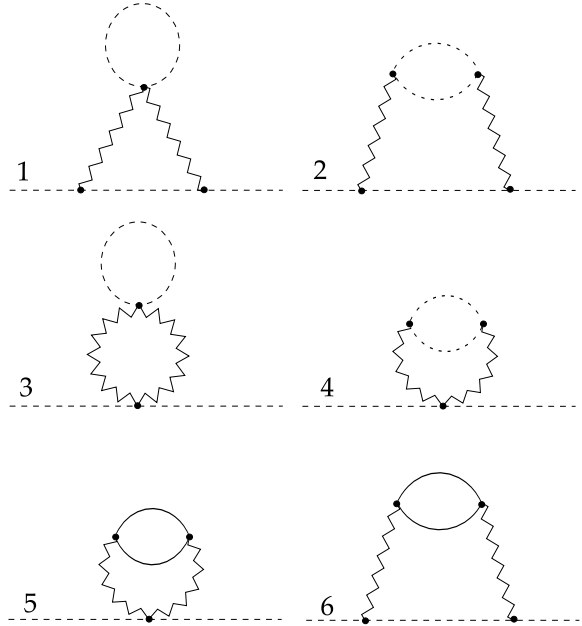}}
 \caption{\footnotesize Graphs corresponding to gauge boson exchange.
 In the minimal gauge mediation case there is just the contribution from a massless gauge boson;
 in our setting the contribution of both the massless and the massive gauge bosons must be taken into account. }
\label{many-graphs}
\end{figure}

The graphs which give the contribution due to the exchange of
gauge bosons are shown in figure \ref{many-graphs}.
In the case of minimal gauge mediation   \cite{Martin1996}, which corresponds to
the $v \rightarrow \infty$  limit of our setting,
only the contribution of a massless   gauge boson must be taken into account.

In our more general setting, we can
introduce the following mass eigenstates:
\beq
A^A_\mu= \frac{g_2 A^1_\mu + g_1 A^2_\mu }{\sqrt{g_1^2+g_2^2}} \, , \qquad
A^B_\mu=  \frac{g_1 A^1_\mu - g_2 A^2_\mu }{\sqrt{g_1^2+g_2^2}} \,  ,
\eeq
The combination $A^A_\mu$ is massless, while $A^B_\mu$ get a mass $m_v$
due to Higgs mechanism.
The covariant derivatives of $Q$ and $T$ in the new variables are:
\beq
D_\mu Q=\partial_\mu Q + ig_e A_\mu^A Q + \frac{i g_1^2}{\sqrt{g_1^2+g_2^2}} A_\mu^B Q \, ,
\eeq
\[
D_\mu T=\partial_\mu T + ig_e A_\mu^A T - \frac{i g_2^2}{\sqrt{g_1^2+g_2^2}} A_\mu^B T \, ,
\]
where $g_e$ is defined in (\ref{gsm}).

Let us denote with $k$ the momentum on the gauge boson propagators.
Three kinds of graphs must then be taken into account: the one with two massless $A^A_\mu$
propagators, the ones with two massive  $A^B_\mu$ exchanges and the ones with one
massless and one massive propagators. The contribution of the last kind of graphs
 comes with a relative minus sign with respect
to the first two; the result is:
\beq
\int\frac{d^4 p}{(2 \pi) ^4}\frac{d^4 k}{(2 \pi)^4}  \left( \cite{Martin1996}\right)
\left( 1+\frac{(k^2)^2}{(k^2-m_v^2)^2}
 - \frac{2 k^2}{(k^2-m_v^2)}
 \right)  \, .
\eeq
Here (\cite{Martin1996}) is the same as the integrand for MGM in \cite{Martin1996},
while the expression in the second parentheses gives the common factor $f(k^2,m_v^2)$,
where the momentum $k$ corresponds to the one on the
massless propagator. This proves the claim in eq. (\ref{mmff}) for the gauge boson graphs;
the detailed evaluation of the graphs
is presented in appendix B.

\subsection{Gaugino  graphs}

The contribution due to gaugino exchange is given
by three classes of graphs, one for the combination of the two gauginos
that is massless at tree level, one for the combination that
gets a tree-level Dirac mass, and a mixed one. 
It is very useful for the evaluation to use the Feynman rules
given in \cite{denner} for Majorana fermions and for
interactions with explicit charge conjugate spinors.

\begin{figure}[h]
 \centerline{\includegraphics[width=5in]{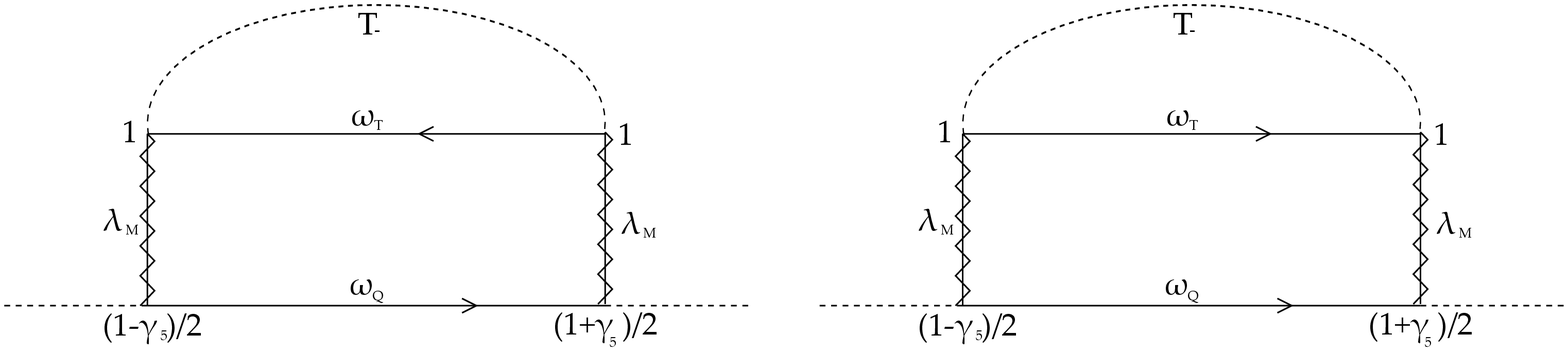}}
 \caption{\footnotesize Contribution due to the mediation
of the massless gaugino.}
\label{gr8-1}
\end{figure}

\begin{figure}[h]
 \centerline{\includegraphics[width=5in]{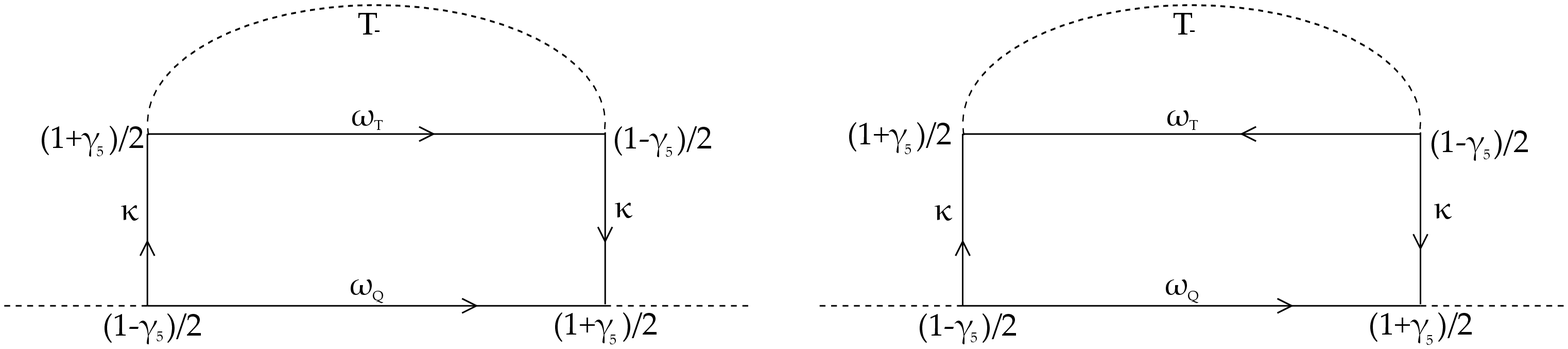}}
 \caption{\footnotesize Contribution due to the mediation
of the dirac fermion $\kappa$.}
\label{gr8-2}
\end{figure}

We first recall the evaluation of the gaugino graph in MGM.
The contribution due to $T_-$  is shown in figure \ref{gr8-1};
the contribution due to $T_+$ is similar~\footnote{There are
some extra $\pm\gamma_5$ factors which
at the end give rise to the same evaluation,
with the replacement $m_- \rightarrow m_+$.}.
The evaluation gives:
\[
4 g^4 \int \frac{d^4 p}{(2 \pi) ^4}  \frac{d^4 k}{(2 \pi)^4}
 \frac{\Tr (\slashed{k} \frac{1-\gamma_5}{2}\slashed{k}
 \frac{1+\gamma_5}{2}\slashed{k}(\slashed{k}-\slashed{p} +m_f)) }
 {(k^2)^3  ((k-p)^2-m_f^2)(p^2-m_\pm^2)}
 \]
\[
= 4 g^4 \int \frac{d^4 p}{(2 \pi) ^4}  \frac{d^4 k}{(2 \pi)^4}
 \frac{2 (k^2 -kp) }
 {(k^2)^2  ((k-p)^2-m_f^2)(p^2-m_\pm^2)}~.
\]
In the case of MgGM there is the same diagram,
corresponding to the exchange of the massless
gaugino $\lambda_A$, weighted by $g^4=g_e^4$.

There is also a diagram corresponding to the exchange
of the Dirac fermion $\kappa$ (see figure \ref{gr8-2}):
\[  4 g_e^4 \,
 \int \frac{d^4 p}{(2 \pi) ^4}  \frac{d^4 k}{(2 \pi)^4}
 \frac{2 k(k-p)}
 {(k^2-m_v^2)^2 ((k-p)^2-m_f^2) (p^2-m_{\pm}^2)} \, .
\]
\begin{figure}[h]
 \centerline{\includegraphics[width=5in]{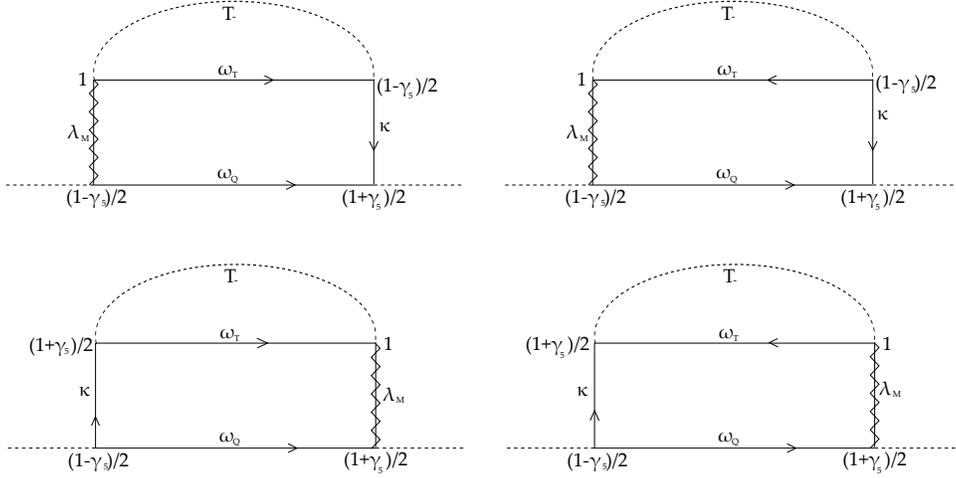}}
 \caption{\footnotesize Mixed contribution
 due to the combined action of $\lambda_M, \kappa$.}
\label{gr8-3}
\end{figure}
Finally, there are also mixed diagrams which exchange both $\lambda_M,\kappa$,
as shown in figure \ref{gr8-3}. They have a minus sign with respect to the previous
ones, and they all give the same contribution; the total is:
\[ -8  g_e^4 \,
 \int \frac{d^4 p}{(2 \pi) ^4}  \frac{d^4 k}{(2 \pi)^4}
 \frac{2k(k-p)}
 {(k^2-m_v^2) (k^2) ((k-p)^2-m_f^2) (p^2-m_{\pm}^2)} \, .
\]
All in all, the sum of the three kinds of diagrams is:~\footnote{More manipulations
with this integral are presented in appendix B.}
\[
4 g_e^4 \int \frac{d^4 p}{(2 \pi) ^4}  \frac{d^4 k}{(2 \pi)^4}
 \frac{2 (k^2 -kp) }
 {(k^2)^2  ((k-p)^2-m_f^2)(p^2-m_\pm^2)}
 \left( 1+\frac{(k^2)^2}{(k^2-m_v^2)^2}
 - \frac{2 k^2}{(k^2-m_v^2)}
 \right) \, ,
\]
which gives the same factor inside the integral as for the
other contributions. This completes the proof of eq. (\ref{mmff}).

\section{Evaluation of the integrals}
\label{integrals-section}

In this section, we write the integrals for the sfermions mass in a notation similar to the one in
\cite{veltman}; we pass to Euclidean variables, and define
\beq
\langle m_{11}, \ldots, m_{1 n_1} | m_{21} , \ldots , m_{2 n_2} | m_{31},
 \ldots, m_{3 n_3} \rangle
\eeq
\[
= \int \frac{d^d k}{\pi^{d/2}}  \frac{d^d q}{\pi^{d/2}} \prod_{i=1}^{n_1}
\prod_{j=1}^{n_2} \prod_{l=1}^{n_3} \frac{1}{k^2+m_{1i}^2}
 \frac{1}{q^2+m_{2j}^2}  \frac{1}{(k-q)^2+m_{3l}^2} \, .
\]
In this notation the integral that should be evaluated in order to compute the sfermions mass is:
\beq
(g_e^4 m_v^4/(4 \pi)^d) \left( - \langle m_+|m_+|0,m_v,m_v \rangle  -  \langle m_-|m_-|0,m_v,m_v  \rangle \right.
\eeq
\[ \left.
-4 \langle m_f|m_f|0,m_v,m_v  \rangle  -2  \langle m_+|m_-|0,m_v,m_v  \rangle + 4  \langle m_+|m_f|0,m_v,m_v  \rangle  \right.
\]
\[
\left.  +4 \langle m_-|m_f|0,m_v,m_v  \rangle -4 m_+^2 \langle m_+|m_+|0,0,m_v,m_v  \rangle \right.
\]
\[ \left. - 4 m_-^2  \langle m_-|m_-|0,0,m_v,m_v  \rangle +  8 m_f^2 \langle m_f|m_f|0,0,m_v,m_v  \rangle \right.
\]
\[ \left. +4(m_+^2 -m_f^2)   \langle m_+|m_f|0,0,m_v,m_v  \rangle  +
 4(m_-^2 -m_f^2)   \langle m_-|m_f|0,0,m_v,m_v  \rangle
\right) \, .
\]
Note that this is obtained from the result in \cite{Martin1996} by adding
the last two entries in each term: $\langle\cite{Martin1996}\rangle\to
\langle \cite{Martin1996},m_v,m_v\rangle$.

We will use the following expression taken from \cite{veltman}, with the convention $d=4-2 \epsilon$:
\beq
 \langle  m_0|m_1|m_2  \rangle=
\frac{1}{-1+2 \epsilon} \left(
m_0^2    \langle  m_0,m_0|m_1|m_2  \rangle  \right.
\eeq
\[ \left. + m_1^2   \langle m_1,m_1|m_0|m_2  \rangle+
m_2^2  \langle  m_2,m_2|m_0|m_1  \rangle
\right) \, .
\]
The basic object to compute  then is
\beq
\langle m_0,m_0|m_1|m_2 \rangle=
\frac{1}{2 \epsilon^2}+\frac{1/2-\gamma - \log m_0^2}{\epsilon}
\eeq
\[
+\gamma^2-\gamma +\frac{\pi^2}{12}
+(2 \gamma -1) \log m_0^2 + \log^2 m_0^2 -\frac{1}{2}+ h(a,b) \, .
\]
The function $h$ is given by the integral  \cite{veltman}:
\beq
h(a,b)= \int_0^1 dx  \left( 1+ {\rm Li}_2 (1-\mu^2) -\frac{\mu^2}{1-\mu^2}  \log \mu^2 \right) \, ,
\eeq
where the dilogarithm is defined by ${\rm Li}_2(x)=-\int_0^1 {dt\over t}\log(1-xt)$,
$a=m_1^2/m_0^2$, $b=m_2^2/m_0^2$,
and
\beq
\mu^2=\frac{a x + b(1-x)}{x(1-x)} \, .
\eeq
For $a=0$, the function $h$
simplifies to  $h(0,b)=1+\rm{Li}_2 (1-b)$.
It is also possible to write an analytical expression:
\beq
h(a,b)=1-\frac{\log a \log b}{2} -\frac{a+b-1}{\sqrt{\Delta}} \left( {\rm Li}_2 \left (-\frac{u_2}{v_1} \right) + {\rm Li}_2 \left (-\frac{v_2}{u_1} \right)
\right.
\eeq
\[
\left. + \frac{1}{4} \log^2 \frac{u_2}{v_1}  +  \frac{1}{4} \log^2 \frac{v_2}{u_1} +    \frac{1}{4} \log^2 \frac{u_1}{v_1} -    \frac{1}{4} \log^2 \frac{u_2}{v_2} + \frac{\pi^2}{6} \right) \, ,
\]
where
\beq
\Delta= 1-2 (a+b) +(a-b)^2 \, ,  \qquad u_{1,2}= \frac{1+b-a \pm \sqrt{\Delta}}{2} \, ,
\eeq
\[ v_{1,2}=\frac{1-b+ a \pm \sqrt{\Delta} }{2} \, .
\]
The integrals with two massless propagators are infrared
divergent and so a mass $m_\epsilon$ must be introduced there as an infrared cutoff;
this artificial parameter will disappear at the end of the calculation.
A useful relation \cite{Martin1996} is:
\beq
\langle m_a | m_b | m_\epsilon, m_\epsilon \rangle = \frac{\Gamma(1+2 \epsilon)}{2}
\left( \frac{1}{\epsilon^2}+\frac{1-2 \log m_\epsilon^2 }{\epsilon} +1 -\frac{\pi^2}{6}  \right.
\eeq
\[ \left. - F_2(m_a^2,m_b^2)- 2 F_3(m_a^2,m_b^2) + (-2 +2 F_1(m_a^2,m_b^2) \log m_\epsilon^2 + \log^2 m_\epsilon^2)
\right) \, ,
\]
where
\beq
F_1(a,b)=\frac{a \log a -b \log b}{a-b} \, , \qquad
F_2(a,b)= \frac{a \log^2 a - b \log^2 b}{a-b } \, ,
\eeq
\[
F_3(a,b)= \frac{a \, {\rm Li}_2 (1-b/a) -b \, {\rm Li}_2(1-a/b) }{a-b} \, ,
\]
for $a \neq b$ and
\beq
F_1(a,a)=1+\log a \, , \qquad F_2(a,a)=2 \log a +\log^2 a \, , \qquad F_3(a,a)=2 \, .
\eeq
We can then use the following expressions \cite{ghinculov} to relate the integrals
to the known objects   $\langle m_0|m_1|m_2 \rangle$ or
$\langle m_0,m_0|m_1|m_2 \rangle$:
\beq
 \langle m_a | m_b | 0, m_v, m_v  \rangle = \frac{  \langle m_a | m_b | 0 \rangle -   \langle m_a | m_b | m_v \rangle }{m_v^4}
-\frac{ \langle m_a | m_b | m_v , m_v  \rangle}{m_v^2}  \, ,
\eeq
\[
\langle  m_a | m_b | m_\epsilon,m_\epsilon, m_v, m_v  \rangle
\]
\[
= \frac{\langle m_a | m_b |  m_v, m_v  \rangle + \langle m_a | m_b | m_\epsilon,m_\epsilon  \rangle }{(m_v^2-m_\epsilon^2)^2}
+2 \frac{ \langle m_a | m_b | m_v \rangle - \langle m_a | m_b | m_\epsilon  \rangle  }{(m_v^2-m_\epsilon^2)^3} \, .
\]
The sfermions mass can be expressed as:~\footnote{The
$4$ factor is due to our choice of $U(1)$ charges.}
\beq
m_{\tilde{f}}^2
=4 \left( \frac{F}{M} \right)^2 \,  \left( \frac{\alpha_e}{4 \pi}  \right)^2 \, s(x,y) \, ,
\eeq
where $x$ and $y$ are defined in eq. (\ref{xxyy}); note that $x<1$ (to avoid unstable messengers).

 \begin{figure}[h]
\begin{center}
$\begin{array}{c@{\hspace{.2in}}c@{\hspace{.2in}}c} \epsfxsize=2.5in
\epsffile{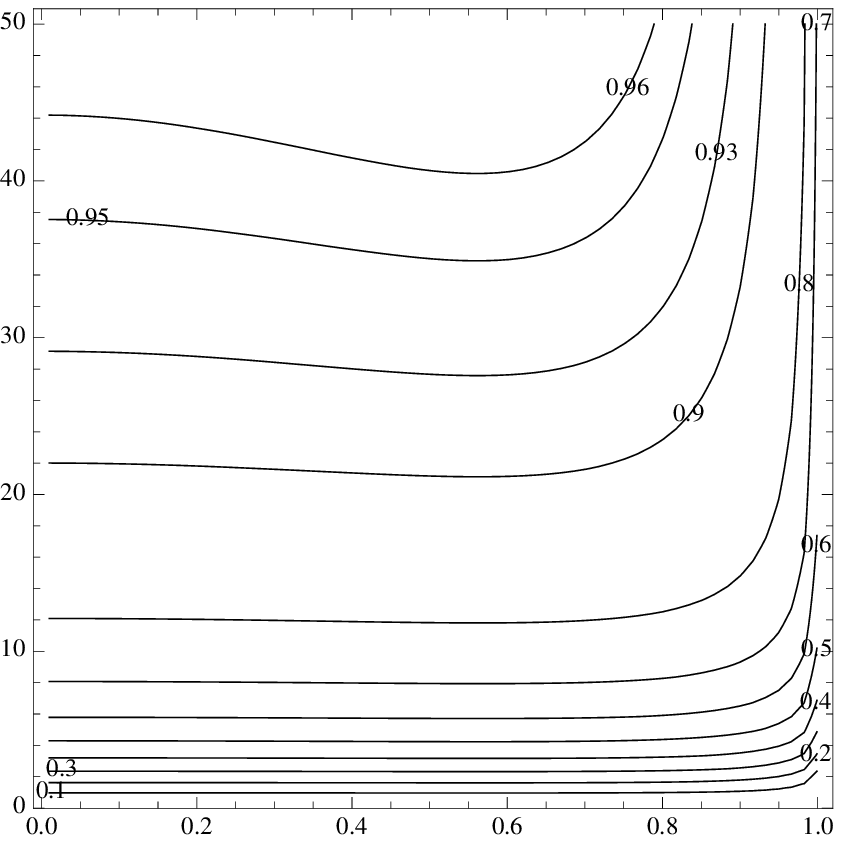}  &
     \epsfxsize=2.5in
    \epsffile{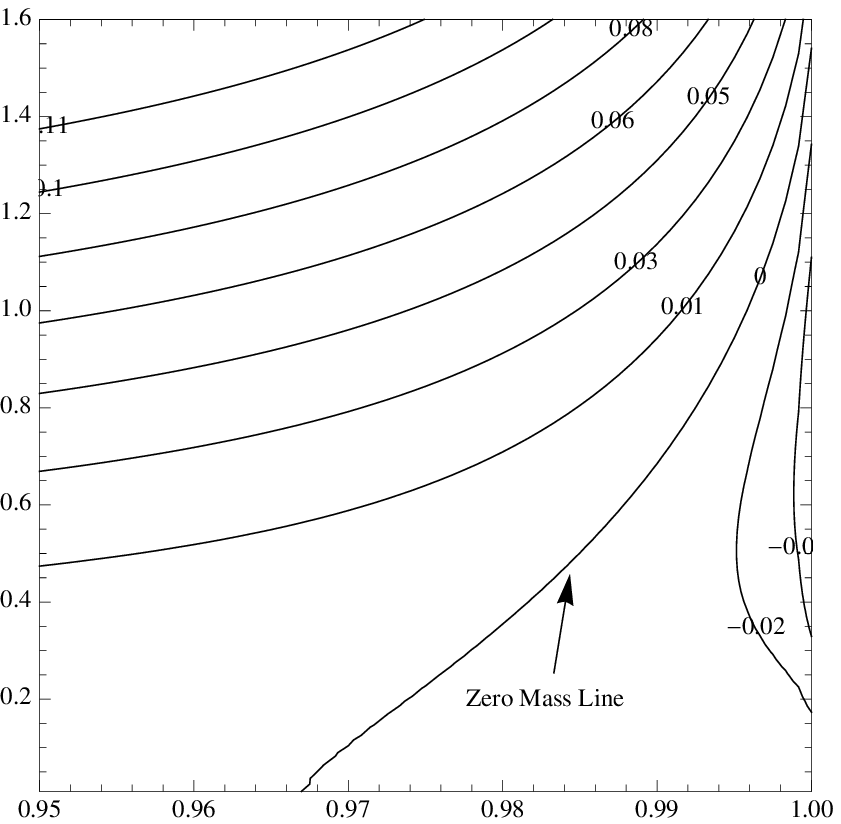}
\end{array}$
\end{center}
\caption{\footnotesize
Contour plot for $s(x,y)$. On the right we zoom on the regime
near $x=1$ and small $y$,
and we find that the sfermion is tachyonic below the
zero mass line.}
\label{ressa0}
\end{figure}

 \begin{figure}[h]
\begin{center}
$\begin{array}{c@{\hspace{.2in}}c@{\hspace{.2in}}c} \epsfxsize=2.5in
\epsffile{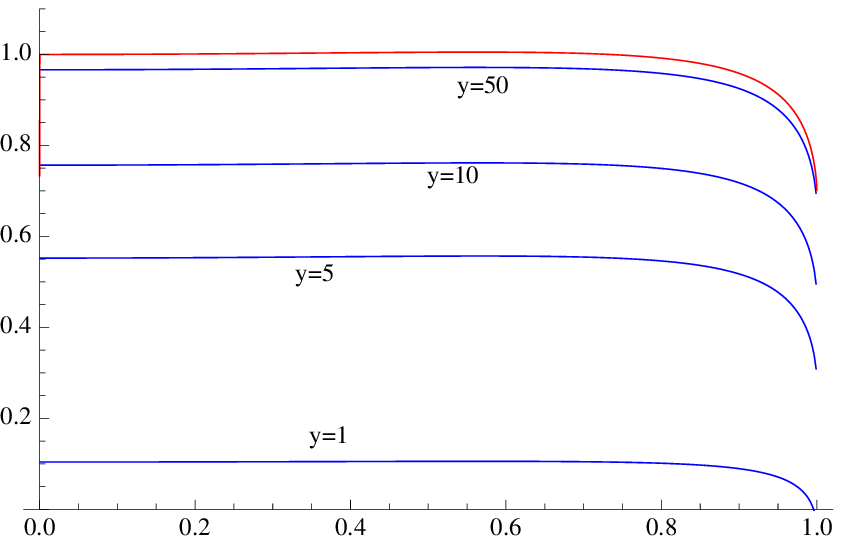}  &
     \epsfxsize=2.5in
    \epsffile{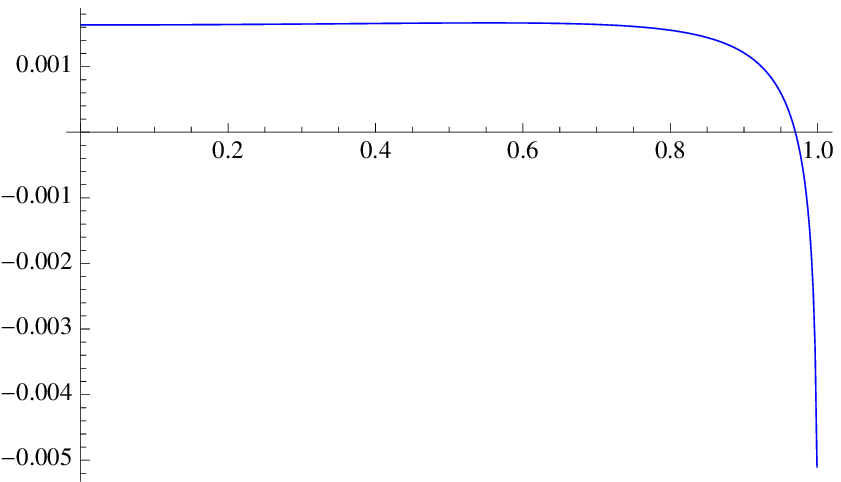}
\end{array}$
\end{center}
\caption{\footnotesize
Left: the function $s(x,y)$, plotted along the $x$ axis for $y=1,5,10,50$.
The top line corresponds to the gauge mediation case
 (formally $y \rightarrow \infty$). Right: the same plot for $y=1/10$.
 The sfermion becomes tachyonic near $x =1$.   }
\label{ressa1}
\end{figure}

 \begin{figure}[h]
\begin{center}
$\begin{array}{c@{\hspace{.2in}}c@{\hspace{.2in}}c} \epsfxsize=2.5in
\epsffile{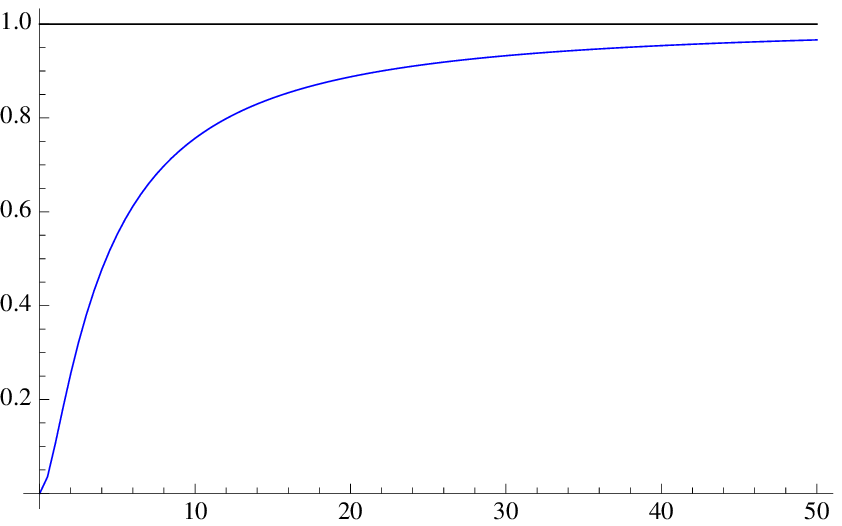}  &
     \epsfxsize=2.5in
    \epsffile{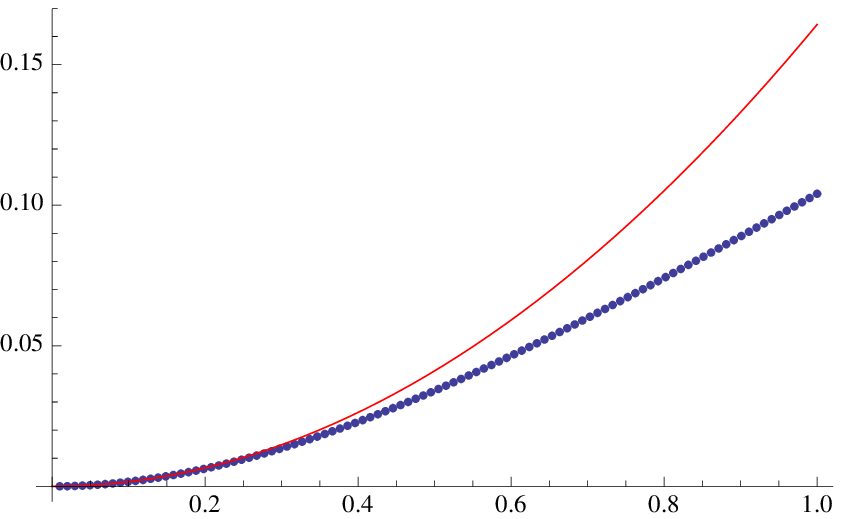}
\end{array}$
\end{center}
\caption{\footnotesize
The function $s(x,y)$, plotted along the $y$ axis for $x=1/100$.
On the right we zoom on the small $y$ regime; the upper line corresponds to a quadratic fit
on the values with $y \leq 0.1$, which gives  $s \approx 0.1643 \, y^2$ in this regime; this
fit is a good approximation as long as $m_v<M/3$. }
\label{ressa2}
\end{figure}

The analytic expression for $s(x,y)$ is:
\beq
s(x,y)={1\over 2x^2}\left(s_0 +\frac{s_1+ s_2}{y^2} + s_3 +s_4 +s_5 \right)
 + \, (x \rightarrow -x)\, , \label{esse}
\eeq
where
\beq
s_0=2(1+x) \left( \log (1+x) -2 {\rm Li}_2  \left( \frac{ x}{1+x}\right)
+\frac{1}{2} {\rm Li}_2 \left( \frac{2 x}{1+x}\right) \right)   \, , \label{mgmmasses2}
\eeq
\[
s_1=- 4 x^2  - 2 x(1+x) \log^2(1+x) - x^2 \, {\rm Li}_2(x^2) \, ,
\]
\[
s_2=8 \left(1+x\right)^2 h\left(\frac{y^2}{1+x},1\right)-4 x \left(1+x\right) h\left(\frac{y^2}{1+x},\frac{1}{1+x}\right)
\]
  \[
   -4 x    h\left(y^2,1+x \right)-8 h\left(y^2,1\right) \, ,
\]
\[
s_3= -2 h\left(\frac{1}{y^2},\frac{1}{y^2}\right)
-2 x \,   h\left(\frac{1+x}{y^2},\frac{1}{y^2}\right) +
2(1+ x) h\left(\frac{1+x}{y^2},\frac{1+x}{y^2}\right) \, ,
\]
\[
s_4=(1+x) \left(  2  h\left(\frac{y^2}{1+x},\frac{1}{1+x}\right)
 - h\left(\frac{y^2}{1+x},1\right)- h\left(\frac{y^2}{1+x},\frac{1-x}{1+x}\right)  \right) \, ,
\]
\[
s_5= 2 h\left(y^2,1+x\right)-2 h\left(y^2,1\right) \, .
\]
The expressions $s_0(x)$ and $s_1(x)$ were simplified by using standard
dilogarithm identities.
Note that in the $y \rightarrow \infty$ limit only $s_0$ contributes;
the result then reduces to the one in minimal gauge mediation \cite{Dimopoulos:1996gy,Martin1996}.

Some plots of the function $s(x,y)$ are shown in figures \ref{ressa0}, \ref{ressa1} and \ref{ressa2}.
In particular, we see that the sfermion is tachyonic in some regime in parameters space.

\section{MSSM sparticle mass spectrum}
\label{mssm-section}

In the case of the MSSM the result for the sfermions mass is:
\beq
m_{\tilde{f}}^2=2 \left( \frac{F}{M} \right)^2 \sum_r  \left( \frac{\alpha_r}{4 \pi} \right)^2
C_r^{\tilde{f}}n_r s(x,y_r) \, , \label{mssmres}
\eeq
where
\beq
y_r=\frac{m_{v_r}}{M}~, \qquad m_{v_r}=2v\sqrt{\left(g_1^{(r)}\right)^2+\left(g_2^{(r)}\right)^2}~,
\eeq
with $g_{1,2}^{(r)}$ being the couplings of $G_{SM_{1,2}}$ in figure \ref{quiver},
respectively; $r=1,2,3$ for $U(1),SU(2),SU(3)$, respectively, and
The corrected version is:
\beq
\alpha_r\equiv{\left(g_{SM}^{(r)}\right)^2\over 4\pi}~,\qquad
\frac{1}{\left(g_{SM}^{(r)}\right)^2}=
\frac{1}{\left(g_1^{(r)}\right)^2 }+\frac{1}{\left(g_2^{(r)}\right)^2 }~.\label{ggrr}
\eeq
In eq. (\ref{mssmres}), $C_r^{\tilde{f}}$ is the quadratic Casimir invariant of
the MSSM scalar field $\tilde{f}$, in a normalization where
$C_3=4/3$ for color triplets, $C_2=3/4$ for $SU(2)$ doublets and
$C_1={3\over 5}Y^2$; $n_r$ is the Dynkin index for the pair of messengers
in a normalization where $n_r=1$  for $N+\bar{N}$ of $SU(N)$, and $n_1={6\over 5}Y^2$
for a messenger pair with weak hypercharge $Y=Q_{\rm EM}-T_3$
(we use the GUT normalization for $\alpha_1$, as in \cite{Martin1996}).

In the limit $m_v \rightarrow \infty$ the well known result of \cite{Dimopoulos:1996gy,Martin1996}
is recovered, with $s=t(x)$ (see the previous section):
\beq
t(x)=\frac{1+x}{x^2} \left( \log (1+x) -2 {\rm Li}_2  \left( \frac{ x}{1+x}\right)
+\frac{1}{2} {\rm Li}_2 \left( \frac{2 x}{1+x}\right) \right) + \, (x \rightarrow -x)  \, . \label{mgmmasses}
\eeq
The gauginos mass is instead the same as in minimal gauge mediation:
\beq
m_{\tilde{g}_r}=\frac{\alpha_r}{4 \pi} \frac{F}{M} n_r \, q(x) \, , \label{mgauginos}
\eeq
where $\alpha_r$ are given in (\ref{ggrr}), and
\beq
q(x)=\frac{1}{x^2} \left( (1+x ) \log(1+x) + (1-x) \log(1-x) \right) \, .  \label{gauginomass}
\eeq

\section{Discussion}
\label{conclusion-section}

In this note we computed the sparticle mass spectrum in Minimal gaugino-Gauge Mediation (MgGM)
as a function of the parameters $x$ and $y$ in (\ref{xxyy}).
We have not studied the Renormalization Group Evolution of the soft masses,
and it should be interesting to investigate how it affects the sparticle
spectrum at the weak scale.

One peculiar result is that in low-scale gaugino mediation,
the sfermions become tachyonic (at the messenger scale $M$)
when the effective SUSY-breaking scale, $F/M$, approaches $M$.
This occurs in a very small corner of the $(x,y)$ plane,
where it is likely that the RGE flips the sign of $m^2_{\tilde{f}}$.
{}For small $v$, there are also important three-loop contributions \cite{gkk},
which we ignored in this note; in particular,
these may also cure the instabilities mentioned above.

The models studied here provide a particular class of General Gauge Mediation (GGM) models
 \cite{GGM} (although they do not fall into the class of
General Messenger Gauge Mediation (GMGM) models \cite{GMGM}).
Some possible generalizations of our work are the following.
First, one may define General gaugino-Gauge Mediation (GgGM) models and compute their soft masses.
In particular, it will be interesting
to compute the soft masses in the ``Direct Gaugino Mediation'' models of \cite{gkk}
and their generalizations, namely, in dynamical realizations of MgGM and its generalizations
in (deformed) SQCD.
It should also be interesting
to find which of the parameters space of GGM is being covered, and to investigate the
phenomenological aspects, e.g. constraints on the spectrum, the NLSP
and the experimental signatures
for the classes of models above.

\bigskip
\noindent{\bf Note Added:}
The result (\ref{factor},\ref{mmff}) was generalized to an arbitrary SUSY-breaking sector
in \cite{Sudano:2010vt}.
See also the recent work \cite{McGarrie:2010qr}.

\bigskip
\noindent{\bf Acknowledgements:}
We are grateful to Zohar Komargodski for fruitful discussions.
A.G. thanks the theory division at CERN for hospitality.
This work was supported in part
by the BSF -- American-Israel Bi-National Science Foundation,
by a center of excellence supported by the Israel Science Foundation
(grant number 1468/06), DIP grant H.52, and the Einstein Center at the Hebrew University.

\section*{Appendix A - One-loop gaugino masses}

For completeness, the 1-loop gaugino mass is presented; it is given
by the MGM one, with the gauge couplings in the unbroken $G_{SM}$ group.
This is obtained from the sum of two diagrams,
one with the scalar messenger with mass $m_-$ (whose coupling
is proportional to $\bar{\lambda} _M \omega_T $)
and one with mass $m_+$  (whose coupling
is proportional to $\bar{\lambda}_M  ( \gamma^5 ) \omega_T $) running in the loop:
\[ g^2
\int   \frac{d^4 k}{(2 \pi)^4}  \frac{F}{(k^2-M^2+F)
(k^2-M^2 -F)} \frac{\slashed{k} + M}{k^2-M^2 }
\]
\[
= g^2 \int_0^1 dx \int_0^{1-x} dy
 \int   \frac{d^4 k}{(2 \pi)^4}  \frac{2 F M}{(k^2 -M^2 -F (y-x) )^3} \, .
\]
Going to Euclidean variables, the evaluation gives:
\[ \frac{\alpha}{4 \pi} \frac{F}{M} \int_0^1 dx \int _0^{1-x} dy \frac{1 }{1+(x-y) \frac{ F}{M^2}} \, , \]
which after an integration gives the well known result
(which is in eqs. (\ref{mgauginos},\ref{gauginomass}) of this note).
In the case of MgGM, the same formula applies
with $g=g_e$ (\ref{gsm}), since the gaugino is in the gauge multiplet of the unbroken $G_{SM}$ group.

\section*{Appendix B - Evaluation of the gauge boson and gaugino graphs}

Let us evaluate the graphs in figure \ref{many-graphs} explicitly; Feynman gauge is used.
The evaluation of graph 1 is:
\beq
2 g_e^4  \int \frac{d^4 p}{(2 \pi) ^4} \frac{1}{p^2-m_{\pm}^2}   \int \frac{d^4 k}{(2 \pi)^4} \frac{1}{(k^2)^2}   \, f(k^2,m_v^2) \, ,
\eeq
where $f(k)$ is given in (\ref{factor}), and
there is a symmetry factor $S=2$.
Here and below, a $\sum_{m_+,m_-}$ is understood.

The evaluation of graph 2 gives:
\[
-  g_e^4 \int \frac{d^4 p}{(2 \pi) ^4}  \frac{d^4 k}{(2 \pi)^4}
\frac{((2 p +k)k)^2}{(k^2)^3 ((p+k)^2-m_\pm^2) (p^2-m_\pm^2)}
  \,f(k^2,m_v^2)
\]
\[ =-  g_e^4
\int \frac{d^4 p}{(2 \pi) ^4}  \frac{d^4 k}{(2 \pi)^4}
\left( \frac{1}{p^2-m_\pm^2} - \frac{1}{(p+k)^2 - m_\pm^2}\right)
\frac{2 p k + k^2}{(k^2)^3 }
 \, f(k^2,m_v^2)
\]
\[ =-  g_e^4
\int \frac{d^4 p}{(2 \pi) ^4}  \frac{d^4 k}{(2 \pi)^4}
\frac{-2 p k}{(k^2)^3 ((p+k)^2 - m_\pm^2) }
  \, f(k^2,m_v^2)
\]
\[ =-  g_e^4
\int \frac{d^4 s}{(2 \pi) ^4}  \frac{d^4 k}{(2 \pi)^4}
\frac{-2 s k+2 k^2}{(k^2)^3 (s^2 - m_\pm^2) }
  \, f(k^2,m_v^2)
\]
\[ =-  g_e^4
\int \frac{d^4 s}{(2 \pi) ^4}  \frac{d^4 k}{(2 \pi)^4}
\frac{2 }{(k^2)^2 (s^2 - m_\pm^2) }
 \, f(k^2,m_v^2) \, ,
\]
where at last we have done the change of variable $s=p+k$.
So we have that the total contribution of graphs 1 and 2 cancels,
as in \cite{Martin1996}.

Graph 3 is very similar to graph 1, up to a numerical constant
and negative relative sign:
there is a $4$ coming for the $g_{\mu \nu} g^{\mu \nu}$,
the symmetry factor is $2$ and there is a $4$ from the two
photon-scalar vortices. At the end the evaluation gives $-4$
times graph 1.

A similar strategy can be employed with graph 4:
\[
g_e^4  \int  \frac{d^4 p}{(2 \pi) ^4}  \frac{d^4 k}{(2 \pi)^4}
  \, f(k^2,m_v^2) \,
\frac{(2 p + k)^2}{(k^2)^2 (p^2-m_\pm^2)((p+k)^2-m_\pm)^2}
\]
\[
=  g_e^4  \int \frac{d^4 p}{(2 \pi) ^4} \frac{d^4 k}{(2 \pi)^4}
   \, f(k^2,m_v^2) \,  \left( \frac{4}{(k^2)^2 (p^2-m_\pm^2)}  - \frac{1}{(k^2)(p^2-m_\pm^2)((p+k)^2-m_\pm^2)}
\right. \]
\[ \left.  +
\frac{4 m_\pm^2}{(k^2)^2 (p^2-m_\pm^2) ((p+k)^2-m_\pm^2)}
-\frac{4 p k +2 k^2}{(k^2)^2 (p^2-m_\pm^2) ((p+k)^2-m_\pm^2)}
\right) \, .
\]
The last term is zero because it is proportional to the integral
\[ \int \frac{d^4 p}{(2 \pi) ^4} \frac{d^4 k}{(2 \pi)^4}
\, f(k^2,m_v^2) \,
\frac{2}{(k^2)^2} \left( \frac{1}{p^2-m_\pm^2} - \frac{1}{(p+k)^2-m_\pm^2}\right)=0\, .
\]
The symmetry factor is $S=2$.

The evaluation of graph 5 is:
\[
- g_e^4 \int \frac{d^4 p}{(2 \pi) ^4}  \frac{d^4 k}{(2 \pi)^4}   \, f(k^2,m_v^2) \,
\frac{\Tr  (\gamma^\mu (\slashed{k}+\slashed{p} +m_f)
\gamma^\rho (\slashed{p}+m_f))g_{\mu \rho} }
{  (k^2)^2 (p^2-m_f^2) ((p+k)^2 -m_f^2)}
\]
\[ = g_e^4
\int \frac{d^4 p}{(2 \pi) ^4}  \frac{d^4 k}{(2 \pi)^4}    \, f(k^2,m_v^2) \,
\frac{  8 p(p+k) -16 m_f^2 }{  (k^2)^2 (p^2-m_f^2) ((p+k)^2 -m_f^2)}
\]
\[ = g_e^4
\int \frac{d^4 p}{(2 \pi) ^4}  \frac{d^4 k}{(2 \pi)^4}    \, f(k^2,m_v^2)\,
  \left( \frac{4}{(k^2)^2 (p^2-m_f^2)}  + \frac{4}{(k^2)^2 ((p+k)^2-m_f^2)}  \right. \]
 \[ \left.
 -\frac{8 m_f^2}{(k^2)^2 ((p+k)^2-m_f^2) (p^2-m_f^2)}
  - \frac{4}{(k^2) ((p+k)^2-m_f^2) (p^2-m_f^2)}
 \right) \, .
\]
There is a symmetry factor $S=2$.

Now let us check that graph 6 is zero (this is just in Feynman gauge,
which is the one used in the calculation):
\[  g_e^4
\int  \frac{d^4 p}{(2 \pi) ^4}  \frac{d^4 k}{(2 \pi)^4}  \, f(k^2,m_v^2) \,
\frac{k_\mu k_\sigma}{(k^2)^3}
\frac{\Tr (\gamma^\mu (\slashed{k} + \slashed{p} +m_f))
\gamma^\sigma (\slashed{p}+m_f) }{((k+p)^2-m_f^2)(p^2-m_f^2)}
\]
\[ =  g_e^4
\int \frac{d^4 p}{(2 \pi) ^4}  \frac{d^4 k}{(2 \pi)^4}  \, f(k^2,m_v^2) \,
4  \frac{2 (p k)^2+(p k) k^2 -k^2 p^2 + k^2 m_f^2}{(k^2)^3 ((k+p)^2-m_f^2)(p^2-m_f^2)} \, .
\]
Let us then subtract from that
\[ g_e^4 \int \frac{d p^4}{(2 \pi) ^4}  \frac{d^4 k}{(2 \pi)^4}  \, f(k^2,m_v^2)\,  \frac{4 k p}{(k^2)^3 (p^2-m_f^2)} \, ,\]
which is clearly zero by symmetry. What is left is:
\[
-4 g_e^4 \int \frac{d^4 p}{(2 \pi) ^4}  \frac{d^4 k}{(2 \pi)^4}  \, f(k^2,m_v^2)  \,
\left(
\frac{1}{(k^2)^2 ((p+k)^2-m_f^2)} + \frac{k p}{(k^2)^3 ((p+k)^2-m_f^2 )}
\right) \, ,
 \]
which vanishes (this can be shown by using the auxiliary variable
$s=p+k$).

Finally, the gaugino graphs give:
\[
4 g_e^4 \int \frac{d^4 p}{(2 \pi) ^4}  \frac{d^4 k}{(2 \pi)^4}
 \frac{\Tr (\slashed{k} \frac{1-\gamma_5}{2}\slashed{k}
 \frac{1+\gamma_5}{2}\slashed{k}(\slashed{k}-\slashed{p} +m_f)) }
 {(k^2)^3  ((k-p)^2-m_f^2)(p^2-m_\pm^2)}f(k)
 \]
\[
= 4 g_e^4 \int \frac{d^4 p}{(2 \pi) ^4}  \frac{d^4 k}{(2 \pi)^4}
 \frac{2 (k^2 -kp) }
 {(k^2)^2  ((k-p)^2-m_f^2)(p^2-m_\pm^2)}f(k)
\]
\[
= 4 g_e^4 \int \frac{d^4 p}{(2 \pi) ^4} \frac{d^4 k}{(2 \pi)^4}
\left( \frac{1}{(k^2)^2 (p^2-m_\pm^2)}
+ \frac{1}{{(k^2)  ((k-p)^2-m_f^2)(p^2-m_\pm^2)}}  \right.
\]
\[ \left. -\frac{1}{(k^2)^2  ((k-p)^2-m_f^2)}
-\frac{(m_\pm^2-m_f^2)}{(k^2)^2  ((k-p)^2-m_f^2)(p^2-m_\pm^2)}
\right)f(k) \, .
\]


\begin{thebibliography}{100}

\bibitem{Cheng}
  H.~C.~Cheng, D.~E.~Kaplan, M.~Schmaltz and W.~Skiba,
  Phys.\ Lett.\  B {\bf 515} (2001) 395
  [arXiv:hep-ph/0106098].

\bibitem{Csaki}
  C.~Csaki, J.~Erlich, C.~Grojean and G.~D.~Kribs,
  Phys.\ Rev.\  D {\bf 65} (2002) 015003
  [arXiv:hep-ph/0106044].

\bibitem{Kaplan:1999ac}
  D.~E.~Kaplan, G.~D.~Kribs and M.~Schmaltz,
  Phys.\ Rev.\  D {\bf 62}, 035010 (2000)
  [arXiv:hep-ph/9911293].

\bibitem{Chacko:1999mi}
  Z.~Chacko, M.~A.~Luty, A.~E.~Nelson and E.~Ponton,
  JHEP {\bf 0001} (2000) 003
  [arXiv:hep-ph/9911323].

\bibitem{gkk}
  D.~Green, A.~Katz and Z.~Komargodski,
  arXiv:1008.2215 [hep-th].

\bibitem{Martin1996}
  S.~P.~Martin,
  Phys.\ Rev.\  D {\bf 55} (1997) 3177
  [arXiv:hep-ph/9608224].

\bibitem{Dimopoulos:1996gy}
  S.~Dimopoulos, G.~F.~Giudice and A.~Pomarol,
  Phys.\ Lett.\  B {\bf 389}, 37 (1996)
  [arXiv:hep-ph/9607225].

\bibitem{denner}
  A.~Denner, H.~Eck, O.~Hahn and J.~Kublbeck,
  Nucl.\ Phys.\  B {\bf 387} (1992) 467.

\bibitem{veltman}
  J.~van der Bij and M.~J.~G.~Veltman,
  Nucl.\ Phys.\  B {\bf 231} (1984) 205.

\bibitem{ghinculov}
  A.~Ghinculov and J.~J.~van der Bij,
  Nucl.\ Phys.\  B {\bf 436} (1995) 30
  [arXiv:hep-ph/9405418].

\bibitem{GGM}
  P.~Meade, N.~Seiberg and D.~Shih,
  Prog.\ Theor.\ Phys.\ Suppl.\  {\bf 177} (2009) 143
  [arXiv:0801.3278 [hep-ph]].

\bibitem{GMGM}
 D.~Marques,
  JHEP {\bf 0903} (2009) 038
  [arXiv:0901.1326 [hep-ph]];
  T.~T.~Dumitrescu, Z.~Komargodski, N.~Seiberg and D.~Shih,
  JHEP {\bf 1005} (2010) 096
  [arXiv:1003.2661 [hep-ph]].

\bibitem{Sudano:2010vt}
 M.~Sudano,
  arXiv:1009.2086 [hep-ph].


\bibitem{McGarrie:2010qr}
   M.~McGarrie,
  arXiv:1009.0012 [hep-ph].
  
  
\end{thebibliography}
\end{document}